 \documentclass[journal]{IEEEtran}
\ifCLASSINFOpdf
\else
   \usepackage[dvips]{graphicx}
\fi
\usepackage{url}
\usepackage{caption}
\usepackage{graphicx}
\usepackage{cite}
\usepackage{amsmath,amssymb,amsfonts}
\usepackage{graphicx}
\usepackage{textcomp}
\usepackage{xcolor}
\usepackage{MnSymbol}
\usepackage{subcaption}
\usepackage{cuted}
\usepackage[linesnumbered,ruled,vlined]{algorithm2e}
\begin{document}

\title{Multi-cell Coordinated Joint Sensing and Communications\\}
\author{\IEEEauthorblockN{Nithin Babu and }
%\IEEEauthorblockA{Dept. EEE, 
%University College London\\ n.babu@ucl.ac.uk}
\and
\IEEEauthorblockN{Christos Masouros}
%\IEEEauthorblockA{Dept. EEE, 
%University College London\\c.masouros@ucl.ac.uk}
%\and
% \author{Nithin Babu, \IEEEmembership{Student Member, IEEE}, Christos Masouros, \IEEEmembership{Senior Member, IEEE}. 
\thanks{N. Babu and C. Masouros are with Dept. of EEE, University College London, London (e-mail: n.babu@ucl.ac.uk, c.masouros@@ucl.ac.uk). This version has been accepted as a paper in 2023 Asilomar Conference on Signals, Systems, and Computers.}}
% }

\markboth{This version has been accepted as a paper in 2023 Asilomar Conference on Signals, Systems, and Computers}
{Shell \MakeLowercase{\textit{et al.}}: Bare Demo of IEEEtran.cls for IEEE Journals}
\maketitle

\begin{abstract}
This paper proposes block-level precoder (BLP) designs for a multi-input single-output (MISO) system that performs joint sensing and communication across multiple cells and users. The Cramer-Rao-Bound for estimating a target's azimuth angle is determined for coordinated beamforming (CBF) and coordinated multi-point (CoMP) scenarios while considering inter-cell communication and sensing links. The formulated optimization problems to minimize the CRB and maximize the minimum-signal-to-interference-plus-noise-ratio (SINR) are non-convex and are represented in the semidefinite relaxed (SDR) form to solve using an alternate optimization algorithm. The proposed solutions show improved performance compared to the baseline scenario that neglects the signal component from neighboring cells.     
\end{abstract}
\IEEEpeerreviewmaketitle
\section{Introduction}
A common property to realize the next-generation wireless network's location-based services, such as connected vehicles and remote healthcare, is the communication network possessing radio sensing capability. %The process of detecting the presence, location, and speed of objects using the signals transmitted or received  by elements of a communication network is referred to as radio frequency (RF) sensing. Rf sensing is more robust to the variation in the atmospheric condition around the target and ensures more privacy than cameras. This is more useful in indoor applications where the localization service by global navigation satellite systems is limited. 
A high-resolution sensing requires large bandwidth and multiple antennas, which are expected to be a part of the 5G advanced and 6G networks. Moreover, the high path loss in the proposed higher-frequency bands reduces the coverage area, demanding small-cell deployments that increase the chances of line-of-sight (LoS) links to the users/targets. Hence, the next-generation mobile communication network, hereafter referred to as an Integrated Sensing and Communication (ISAC) system, has the potential to do radio frequency (RF) sensing in addition to serving the users.   

The idea of an ISAC system has gained much attention lately from academia \cite{liu2020joint} and industry \cite{wild2021joint}. The main challenge in realizing an ISAC system is designing an optimal waveform tailored to both the sensing and communication performance matrices. Numerous works realize an ISAC system by incorporating communication (radar) information into the existing radar (communication) waveforms \cite{4337601}, \cite{5776640}. Another approach in designing ISAC waveform is to minimize the error with some ideal radar waveform that guarantees a good estimation performance while guaranteeing a set of communication-related constraints \cite{9124713}. Since such solutions depend on the availability of the ideal radar waveform, the authors of \cite{liu2021cramer} proposed optimal waveforms for point and extended targets that minimize the Cramer-Rao Bound (CRB) while guaranteeing a minimum level of signal-to-interference-plus-noise ratio (SINR) for each user. %The CRB determines a lower bound for the variance of any unbiased estimator for parameter estimation. 
Other metrics that have been optimized in the context of an ISAC system include SINR, Mutual information (MI), Energy Efficiency, etc. SINR-based optimization takes the SINR of the radar/communication receivers as the primary objective or constraint. The authors of \cite{7953658} design the radar transmit precoder, the radar subsampling scheme, and the communication transmit covariance matrix to maximize the radar SINR while meeting given communication rate and power constraints. %The MI measures the amount of information conveyed to a communication receiver given the channel state information, whereas, in the radar context, it represents the amount of target information a radar receiver can extract from the channel given the transmitted signal.
A weighted sum of communication and radar MI maximization is presented in \cite{9303435}. The work in \cite{9385108} proposes optimum power allocation schemes to maximize the sum-rate and energy efficiency of an ISAC system while satisfying certain radar target detection and minimum data rate per user requirements.

All the aforementioned works consider a single-cell ISAC system with single/multiple targets and single/multiple users. In practise, there will be more than one ISAC base station (BS) close to each other due to the dense deployment of small cell BSs. This will cause inter-cell interference to the communication users from the neighboring BSs. %, tackled through various resource allocation and precoder design schemes as explained in \cite{8187586}. 
Additionally, inter-cell reflection (ICR) will be received by a BS from its target due to the signal transmitted from the neighboring BSs. The received power through ICR can degrade the target parameter estimation if the BS is unaware of the data transmitted from the neighboring BS. Conversely, the BSs can coordinate by sharing the data to improve the estimation performance. In this paper, we consider a multi-cell ISAC system and design precoders that minimize the CRB in target angle estimation and maximize the minimum SINR received by users, subject to total power constraint. %Our main contributions are
% \begin{itemize}
%     \item We derive the expression to determine the CRB in estimating the azimuth angle of a target with respect to the base station in both the coordinated beamforming and coordinated multi-point transmission schemes.
%     \item These are then utilized to formulate an optimization problem to minimize the CRB value and maximize the minimum SINR value among the communication users subject to minimum SINR and total power constraints. 
%     \item The optimal precoders to be used in the CBF scheme is determined using an alternating optimization method, whereas in the CoMP case, the problem is recast as a convex optimization problem and solved using available solvers.    
% \end{itemize}
\section{System Model}
\begin{figure}[]
\centering
\captionsetup{justification=centering}
\centerline{\includegraphics[width=0.8\columnwidth]{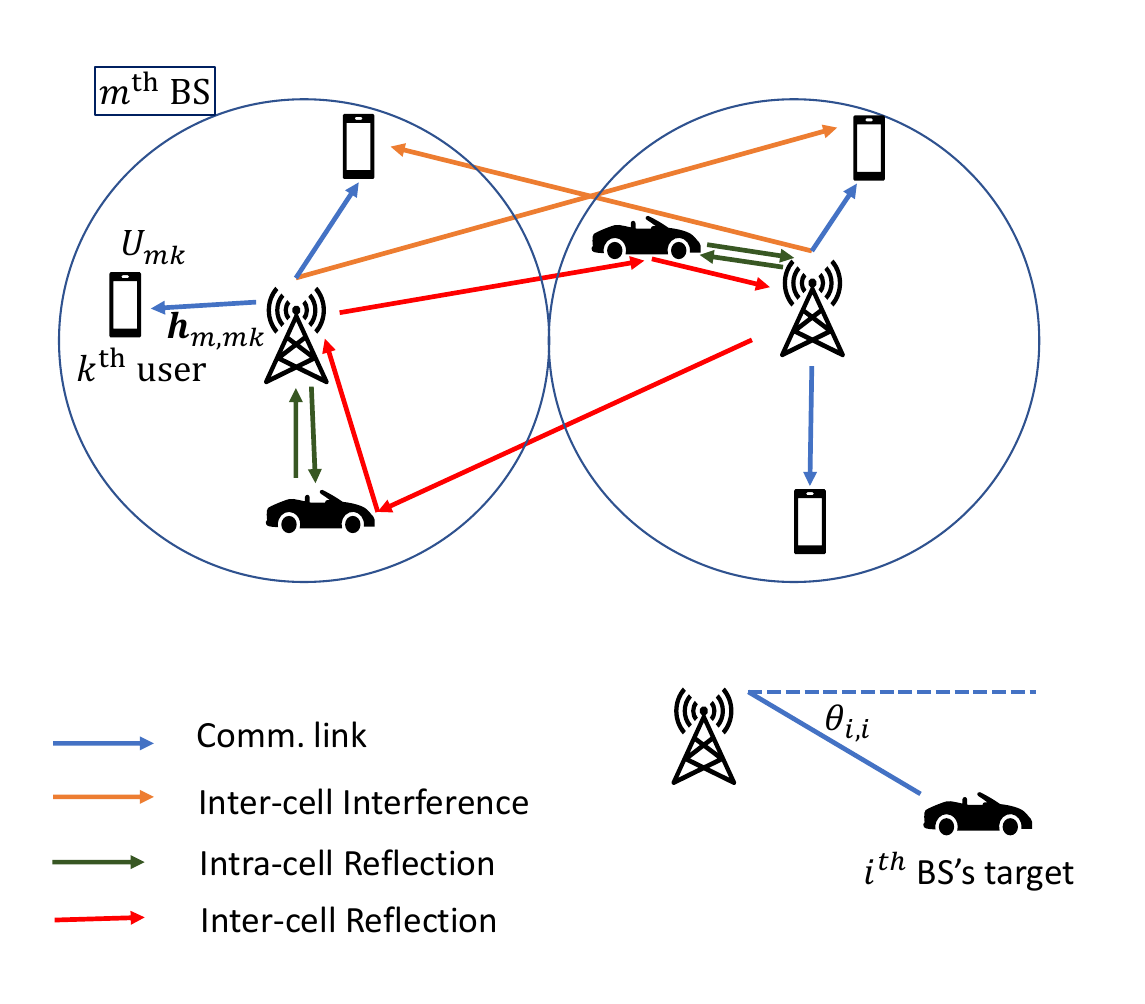}}
\caption{System setup.}
\label{figure1}
\end{figure}
% We consider a multi-cell, multi-user MISO system during downlink transmission. The system consists of $J$ cells, each containing a target, $K$ users and a base station (BS) equipped with $N_{\mathrm{t}}$ transmit and $N_{\mathrm{r}}$ receive antennas. Let $\mathbf{x}_{m}=\sum_{n=1}^{K}\mathbf{w}_{m,n}s_{m,n}\in \mathbb{C}^{N_{\mathrm{t}}\times 1}$ be a narrowband signal vector transmitted by the $m^{\text{th}}$ BS; $\mathbf{w}_{m,n}\in \mathbb{C}^{N_{\mathrm{t}}\times 1}$ and $s_{m,n}$ denote the precoding and transmitted data to $n^{\text{th}}$ user covered by the $m^{\text{th}}$ BS: $U_{m,n}$. Since all the BSs transmit the respective signal vector, $\{\mathbf{x}_{m}\}$, simultaneously, the received reflected echo signal at the $m^{\text{th}}$ BS from the target in the cell is given by,
%\color{blue}To enable low-complexity and energy-efficient receivers, we assume single-user detection meaning that a user is not attempting to decode and subtract interfering signals while decoding its own signals. When the radar has no a priori knowledge about targets,
% %the initial step is to search for potential targets in the whole
% space. Similarly, when no channel information is available
% at the communication system, the CSI has to be estimated
% before any useful information can be decoded at the receiver. More specifically, in our case, the BS first sends omnidirectional DL pilots (DP), and then estimates the K AoAs in  as well as the associated range and Doppler
% parameters of all K targets. 
\color{black}We consider a multi-input single-output (MISO) system with $J$ cells and $K$ users per cell; each cell has a target and a BS with a uniform linear array (ULA) of $N_{\mathrm{t}}$ transmit antennas spaced at $\lambda/2$ distance, where $\lambda$ is the wavelength. The BS is also equipped with a $N_{\mathrm{r}}$-element receive ULA with a sparse spacing of $N_{\mathrm{t}}\lambda/2$ antennas. The $m^{\text{th}}$ BS transmits a narrowband signal matrix, $\mathbf{X}_{m}\in \mathcal{C}^{N_{\mathrm{t}}\times L}$, to the users in the cell, with $L > N_{\mathrm{t}} $ being the length of the radar pulse/ communication frame. Here, all BSs transmit simultaneously, and a BS receives its echo signal and multiple echo signals from its target due to ICR from the neighboring BSs. As shown in the figure, we select the dominant path among the ICR paths. The resulting echo signal received by the $m^{\text{th}}$ BS from the target in its cell is given as
\begin{IEEEeqnarray}{rCl}
\mathbf{Y}_{m}^{\mathrm{R}} &=& \mathbf{G}_{mm}\mathbf{X}_{m}+\sum_{n\neq m}^{J} \mathbf{G}_{nm}\mathbf{X}_{n}+ \mathbf{Z}_{m}^{\mathrm{R}},\quad \forall m,\label{yr} 
%&=& \mathbf{G}_{m} \mathbf{x} + \mathbf{z}_{m}^{\mathrm{R}}.
\end{IEEEeqnarray}
where %$\mathbf{G}_{m}=[\mathbf{G}_{1m}, \mathbf{G}_{2,m},...,\mathbf{G}_{J,m}] \in \mathbb{C}^{N_{\mathrm{r}}\times N_{\mathrm{t}}\cdot J }$ and $\mathbf{x} \in \mathbb{C}^{N_{\mathrm{t}}\cdot J\times 1 }$ .
$\mathbf{G}_{nm}=\alpha_{nm}\mathbf{a}_{mm}\mathbf{v}^{\mathrm{T}}_{nm}\, \forall n = \{1,2,..,J\} \equiv \mathcal{J}$, is the target response matrix at the $m^{\text{th}}$ BS due to the transmission from the $n^{\text{th}}$ BS in which $\mathbf{a}_{m,m}$ and $\mathbf{v}_{nm}$ are the array response vectors in the directions $\theta_{mm}$ and $\theta_{nm}$, respectively, and $()^{\mathrm{T}}$ represents the transpose operation. $\alpha_{n,m}$ represents the complex amplitude of the received signal and $\mathbf{Z}_{m}^{\mathrm{R}}\in \mathbb{C}^{N_{\mathrm{r}}\times L}$ is an additive white Gaussian noise (AWGN) vector with the variance of each entry being $\sigma^{2}_{\mathrm{R}}$. \eqref{yr} assumes that all the neighboring BSs have a LoS link to the $m^{\text{th}}$ BS's target. %Please note that the equation can be easily adapted for a case with a subset of BSs having LoS links with the $m^{\text{th}}$ BS's target. 
The transmitted symbol matrix $\mathbf{X}_{m}=\mathbf{W}_{m}\mathbf{S}_{m}$, where $\mathbf{W}_{m}=
\left[\mathbf{w}_{m1},\mathbf{w}_{m2},...,\mathbf{w}_{mK}\right]$ for $m \in \{1,2,..,J\}$ are  the  dual-functional  beamforming  matrices to be designed. $\mathbf{S}_{m}\in \mathcal{C}^{K \times L}$ is the orthogonal data stream transmitted to $K$ users of the $m^{\text{th}}$ BS: $(1/L)\mathbf{S}_{m}\mathbf{S}^{H}_{m}=\mathbf{I}_{K}$. The first term on the right-hand side (RHS) of \eqref{yr} is called intra-cell reflection due to the signal vector from the same BS, whereas the second term represents ICR due to the signals from the remaining BSs.

The received signal at $k^{\text{th}}$ user of the $m^{\text{th}}$ cell, represented as $U_{mk}$, is expressed as 
\begin{IEEEeqnarray}{rCl}
\mathbf{y}_{mk}^{\mathrm{C}} &=& \mathbf{h}^{T}_{m,mk}\mathbf{X}_{m}+\sum_{n\neq m}^{J} \mathbf{h}^{T}_{n,mk}\mathbf{X}_{n}+ \mathbf{z}_{mk}^{\mathrm{C}}, \, \forall U_{mk},\label{yc} 
\end{IEEEeqnarray}
where $\mathbf{h}_{n,mk}\in \mathbb{C}^{N_{\mathrm{t}}\times 1}$ is the channel from the $n^{\text{th}}$ BS to $U_{m,k}$ and $\mathbf{z}_{mk}^{\mathrm{C}}$ is an AWGN noise vector with variance of each entry being $\sigma^{2}_{\mathrm{C}}$. The first term in the RHS of \eqref{yc} contains intra-cell interference from the users of the same cell, whereas the second term represents the inter-cell interference from the neighboring cells.

 As explained in \cite{bjornson2013optimal}, a multi-cell system can work either in (a) coordinated beamforming (CBF) mode: each BS has a disjoint set of users to serve with data but selects transmit strategies jointly with all other BSs to reduce inter-cell interference, or (b) coordinated multipoint (CoMP) mode in which all the BSs can serve and coordinate interference to all users. In the following section, we consider the CBF and CoMP modes of BS operations to design an efficient precoder that maximizes both the sensing and communication performances. 
% \section{Precoder Design}
% Let us consider the system consists of 2 BSs. Then the echo signal received at BS1 and BS2 are defined as follows 
% \begin{IEEEeqnarray}{rCl}
% \mathbf{Y}_{1} &=& \mathbf{A}_{1,1}\mathbf{X}_{1}+ \mathbf{A}_{2,1}\mathbf{X}_{2}+ \mathbf{Z}_{1}^{\mathrm{R}}\label{Y1} \\
% \mathbf{Y}_{2} &=& \mathbf{A}_{2,2}\mathbf{X}_{2}+ \mathbf{A}_{1,2}\mathbf{X}_{1}+ \mathbf{Z}_{2}^{\mathrm{R}}\label{Y2} 
% %&=& \mathbf{G}_{m} \mathbf{x} + \mathbf{z}_{m}^{\mathrm{R}}.
% \end{IEEEeqnarray}
% $\mathbf{A}_{i,i}= \alpha_{i,i} \mathbf{a}(\theta_{i,i})\mathbf{v}^T(\theta_{i,i})$ for $i \in \{1,2\}$, $\mathbf{A}_{2,1}= \alpha_{2,1}\mathbf{a}(\theta_{1,1})\mathbf{v}^T(\theta_{2,1})$, $\mathbf{A}_{1,2}= \alpha_{1,2}\mathbf{a}(\theta_{2,2})\mathbf{v}^T(\theta_{1,2})$, and $\mathbf{X}_{i}=\mathbf{W}_{i}\mathbf{S}_{i} \in \mathcal{C}^{N_{i} \times L}$, where $\mathbf{W}_{i}=
% \left[\mathbf{w}_{i,1},\mathbf{w}_{i,2},...,\mathbf{w}_{i,K}\right]$ for $i \in \{1,2\}$ are  the  dual-functional  beamforming  matrices  to be designed. $\mathbf{S}_{i}\in \mathcal{C}^{K \times L}$ is the orthogonal data stream transmitted to $K$ users of the $i^{\text{th}}$ BS. Let us assume the parameters to be estimated are the angles of the targets in the cells denoted as $\theta_{1,1}$ and $\theta_{2,2}$. $\mathbf{Z}_{i}^{\mathrm{R}} \in\mathcal{C}^{N_r \times L} $ denotes an additive white Gaussian noise (AWGN)  matrix,  with  the variance  of  each  entry being $P_n$. 
\subsection{Sensing and Communication Performance Metrics}
We aim to minimize the variance  of the error in target parameter estimation. For an unbiased estimator, the error variance is lower bound by the CRB given by the inverse of the Fisher information matrix. We assume each BS estimates its target's angle $\theta_{m,m}$ from the received signal \eqref{yr}. From \eqref{yr}, the received echo signal at the BS is a multi-variate Gaussian random variable with mean $\mathbb{\mu}_{m,*}$ and covariance matrix $\mathbf{C}_{m,*}$. Then the $(m,m)^{\text{th}}$ element of the Fisher information matrix (FIM) is given by,
\begin{align}
\text{F}_{mm} &= 2 \,\text{Re}\,\left\lbrace \text{tr}\left( \frac{d \mathbf{\mu}^{H}_{m,*}}{d \mathbf{\theta}_{mm}} \mathbf{C}_{m,*}^{-1} \frac{d \mathbf{\mu}_{m,*}}{d \mathbf{\theta}_{mm}} \right)\right\rbrace. \label{fim}
\end{align}
 Since we consider only one target parameter estimation per BS, the FIM will be a scalar value given by \eqref{fim}. The entries of the $\mathbf{\mu}_{m,*}$  and $\mathbf{C}_{m,*}$ depend on whether the BSs are operating in the CBF or the CoMP mode, whose corresponding expressions are derived in the following sections.

In the communication aspect, we aim to maximize the minimum signal-to-interference-plus-noise-ratio (SINR) value experienced by a user in a cell. Again, the corresponding SINR expressions vary depending on the BSs' operation mode. 
\subsection{Coordinated Beamforming} \label{seccbf}
In CBF, as no data is shared between the BSs, the $m^{\text{th}}$ BS knows only the data symbol matrix $\mathbf{X}_{m}$. Therefore, from \eqref{yr}, $\boldsymbol{\mu}_{m,\text{cbf}}=\mathbf{G}_{m,m}\mathbf{X}_{m}$ and 
\begin{align}
\mathbf{C}_{m,\text{cbf}} &= L\sum_{n\neq m}^{J} \mathbf{G}_{n,m}\mathbf{W}_{m} \mathbf{W}^{H}_{m}\mathbf{G}^{H}_{n,m} +  \sigma^{2}_{\mathrm{R}}\mathbf{I}_{N_{r}} \label{cmcbf}
%\mathbf{C}_{m} &= \mathbf{A}_{j,i} \mathbf{W}_{j} \mathbf{W}^{H}_{j} \mathbf{A}^{H}_{j,i} +  \sigma^{2}_{\mathrm{R}}\mathbf{I}_{N^{r}_{m}} \quad \{i,j\} \in \{1,2\}, i\neq j
\end{align}

% The covariance matrix of $X$ can be expressed as:

% \begin{align*}
% \mathrm{cov}(X) &= \mathrm{E}[XX^\mathrm{H}] \\
% &= \mathrm{E}[(GWSc)(GWSc)^\mathrm{H}] \\
% %&= \mathrm{E}[GWW^\mathrm{H}G^\mathrm{H}]\mathrm{E}[ScS^\mathrm{H}] \\
% &= GW\mathrm{E}[SS^\mathrm{H}]W^\mathrm{H}G^\mathrm{H} \\
% &= GWW^\mathrm{H}G^\mathrm{H},
% \end{align*}
% where we have used the fact that $\mathrm{E}[SS^\mathrm{H}] = \mathrm{E}[(\mathrm{sqrt}(L)\mathrm{orth}(S^\mathrm{H})^\mathrm{H})(\mathrm{sqrt}(L)\mathrm{orth}(S^\mathrm{H})^\mathrm{H})^\mathrm{H}] = \mathrm{I}_K$ because $\mathrm{orth}(S^\mathrm{H})$ is an orthonormal matrix.

% The covariance matrix of $\mathbf{Z}$ can be expressed as:

% \begin{align}
% \mathrm{cov}(Z) &= \mathrm{E}[ZZ^\mathrm{H}] \\
% &= P\mathbf{I}_N.
% \end{align}

% Therefore, the interference plus the noise term is a multi-variate Gaussian random variable with mean zero and covariance,
% \begin{align}
% \mathbf{Q} &= \mathbf{A}_{2,1} \mathbf{W}_{2,1} \mathbf{W}^{H}_{2,1} \mathbf{A}^{H}_{2,1} + P\mathbf{I}_N.
% \end{align}
% Hence the Fisher Information Matrix with respect to $\theta_{m,m}$ can be estimated as,
% \begin{align}
% \text{F}_{m,m} &= 2 \,\text{RE}\, \text{tr}\left\lbrace \frac{d \mathbf{\mu_{m}}^{H}}{d \mathbf{\theta}_{m,m}} \mathbf{C}_{m}^{-1} \frac{d \mathbf{\mu_{m}}}{d \mathbf{\theta}_{m,m}} \right\rbrace. \label{fim}
% \end{align}
Using the definitions of $\mathbf{G}_{m,m}$, we have,
\begin{align}
&\frac{1}{L\alpha^{2}_{mm}}\text{tr}\left( \frac{d\boldsymbol{\mu}^{H}_{m,\text{cbf}}}{d \mathbf{\theta}_{mm}} \mathbf{C}_{m,\text{cbf}}^{-1} \frac{d \boldsymbol{\mu}_{m,\text{cbf}}}{d \mathbf{\theta}_{mm}} \right)\nonumber\\
&= (\dot{\mathbf{a}}^{H}_{mm} \mathbf{C}_{m,\text{cbf}}^{-1} \dot{\mathbf{a}}_{mm}) \circ  (\mathbf{v}^H_{mm}\mathbf{R}^{*}_{\mathbf{X}_m}\mathbf{v}_{mm})\nonumber\\ 
&+ (\dot{\mathbf{a}}^{H}_{mm} \mathbf{C}_{m,\text{cbf}}^{-1} {\mathbf{a}}_{mm}) \circ  (\mathbf{v}^H_{mm}\mathbf{R}^{*}_{\mathbf{X}_m}\dot{\mathbf{v}}_{mm})\nonumber\\
&+ ({\mathbf{a}}^{H}_{mm} \mathbf{C}_{m,\text{cbf}}^{-1} \dot{\mathbf{a}}_{mm}) \circ  (\dot{\mathbf{v}}^H_{mm}\mathbf{R}^{*}_{\mathbf{X}_m}\mathbf{v}_{mm})\nonumber\\    
&+({\mathbf{a}}^{H}_{mm} \mathbf{C}_{m,\text{cbf}}^{-1} {\mathbf{a}}_{mm}) \circ  (\dot{\mathbf{v}}^H_{mm}\mathbf{R}^{*}_{\mathbf{X}_m}\dot{\mathbf{v}}_{mm})\label{fim_part}
\end{align}
where $\mathbf{R}_{\mathbf{X}_m}= \mathbf{W}_{m}  \mathbf{W}^{H}_{m}=\sum_{k=1}^{K}\mathbf{w}_{mk}  \mathbf{w}^{H}_{mk}=\sum_{k=1}^{K}\mathbf{W}_{mk}$, and $()^{*}$ represents the conjugate of the operand.
% Hence, the CRB in estimating the $\theta_{m,m}$ is given by 
% \begin{align}
% \text{CRB}_{m} &= \frac{1}{2\text{Re}\left\lbrace\frac{1}{L\alpha^{2}_{m,m}}\text{tr}\left( \frac{d \mathbf{\mu}^{H}}{d \mathbf{\theta}_{m,m}} \mathbf{C}_{m,\text{cbf}}^{-1} \frac{d \mathbf{\mu}}{d \mathbf{\theta}_{m,m}} \right)\right\rbrace } \label{crbm}
% \end{align}
Since we aim to estimate one target parameter per BS, minimizing the CRB is the same as maximizing the Fisher information value $\text{FIM}^{\text{cbf}}_{mm}$ obtained by substituting \eqref{fim_part} in \eqref{fim}. Hence, in the CBF mode, for the $m^{\text{th}}$ BS, We aim to solve the following optimization problem:
% \begin{IEEEeqnarray}{rCl}
\begin{align}
&\text{(P1):} \underset{\{\mathbf{w}_{mk}\},\gamma}{\text{maximize}}\,\,\,\,  \frac{u}{\mathrm{NF}_{\mathrm{R}}} \,\, {\text{FIM}^{\text{cbf}}_{mm}} +\frac{(1-u)}{\mathrm{NF}_{\mathrm{C}}} \gamma\nonumber\\
% && \text{F}_{i,i} \geq f \quad \forall i\\
& \frac{|\mathbf{h}^{T}_{m,mk}\mathbf{w}_{mk}|^{2}}{\sum_{l\neq k}^{K}|\mathbf{h}^{T}_{m,mk}\mathbf{w}_{ml}|^{2}+\sum_{n\neq m}^{J}\sum_{l=1}^{K}|\mathbf{h}^{T}_{n,mk}\mathbf{w}_{nl}|^{2}+\sigma^{2}_{\mathrm{C}}} \geq \gamma\IEEEyessubnumber\label{p1.sinr}\\
& \sum_{k=1}^{K}\text{tr}\left ( \mathbf{w}_{mk} \mathbf{w}^{H}_{mk} \right ) \leq P_t, \label{p1.pt}
% & \gamma \geq \Gamma, \label{p1.c3}
\end{align}
where $u$ is the weighting factor to select between the communication and the sensing performance matrices; $\mathrm{NF}_{\mathrm{R}}$ and $\mathrm{NF}_{\mathrm{C}}$ are the normalization factors which are obtained by setting $u=1$ and $u=0$, respectively. The objective function of (P1) is the maximization (minimization) of the FIM (CRB) and the minimum SINR value among the users of the $m^{\text{th}}$ cell. \eqref{p1.sinr} is the SINR constraint whereas, \eqref{p1.pt} is the total power constraint with $P_{t}$ being the total available power at the BS. (P1) is difficult to solve because of the non-convex form of $\mathbf{C}_{m,\text{cbf}}$ in \eqref{fim} and the non-convex multiplication between $\gamma$  and the interference terms in \eqref{p1.sinr}. Using the semidefinite relaxation (SDR) technique, the SINR constraint can be rewritten as, 
\begin{align}
  &  \text{tr}\left (  \mathbf{Q}_{m,mk}  \mathbf{W}_{mk} \right )-\gamma\left(\sum_{l\neq k}^{K} \text{tr}\left (  \mathbf{Q}_{m,mk}  \mathbf{W}_{ml} \right )\right)\nonumber\\
&- \gamma\left(\sum_{n\neq m}^{J}\sum_{l=1}^{K}\text{tr}\left (  \mathbf{Q}_{n,mk}  \mathbf{W}_{nl} \right )\right) \geq \gamma \sigma^{2}_{R} \,\, \forall U_{mk}\label{p1.c1.1},
\end{align}
where $\mathbf{Q}_{m,mk}=\mathbf{h}^{*}_{m,mk}\mathbf{h}^{T}_{m,mk}$ and $\mathbf{W}_{ml}= \mathbf{w}_{ml}\mathbf{w}^{H}_{ml}$. (P1) can be reformulated as,
\begin{align}
&\text{(P1.1):} \underset{\{\mathbf{W}_{mk}\},\gamma}{\text{maximize}}\,\,\,\,  \frac{u}{\mathrm{NF}_{\mathrm{R}}} \,\,{\text{FIM}^{\text{cbf}}_{mm}} +\frac{(1-u)}{\mathrm{NF}_{\mathrm{C}}} \gamma\nonumber,\\
% && \text{F}_{i,i} \geq f \quad \forall i\\
&\eqref{p1.c1.1},\\
& \sum_{k=1}^{K}\text{tr}\left ( \mathbf{W}_{mk} \right ) \leq P_t. \label{p1.c2.1}
\end{align}
% \end{IEEEeqnarray}
% A workaround to the non-convex objective function and the SINR constraint is to solve (P1.1) for the $m^{\text{th}}$ BS while keeping the precoding matrices of all the other BSs fixed and rewrite the SINR constraints as the following signal leakage constraints. This makes the objective function an affine function of $\{\mathbf{W}_{m,k}\}$ through $\mathbf{R}_{x_{m}}=\sum_{k=1}^{K}\mathbf{W}_{m,k}$. The SINR constraint becomes, 
A workaround to the non-convex  SINR constraint is to rewrite it as signal leakage constraints:
\begin{align}
&\sum_{l=1}^{K}\text{tr}\left (  \mathbf{Q}_{m,nk}  \mathbf{W}_{ml} \right ) \leq I_{\text{max}}/(J-1) \,\, \forall U_{n,k} \,\,\forall n \neq m,  \label{intercell}\\
&\sum_{l\neq k}^{K} \text{tr}\left (\mathbf{Q}_{m,mk} \mathbf{W}_{ml}\right ) \leq I^{'}_{\text{max}} \,\, \forall U_{m,k},\label{intracell}\\
&\text{tr}\left (  \mathbf{Q}_{m,mk}  \mathbf{W}_{mk} \right )-\gamma\left(I^{'}_{\text{max}} + I_{\text{max}} \right) \geq \gamma \sigma^{2}_{R} \, \forall U_{mk} \label{sinr}. 
\end{align}
\eqref{intercell} limits the total inter-cell interference experienced by any user to be less than $I_{\text{max}}$ whereas the intracell interference caused is constrained below $I^{'}_{\text{max}}$. Please note that for given $I_{\text{max}}$ and $I^{'}_{\text{max}}$ values, \eqref{intercell}-\eqref{sinr} are all convex constraints of the optimization variable of (P1.1). % The remaining challenge is determining the value of $I_{\text{max}}$ and $I^{'}_{\text{max}}$. From a communication perspective, we would want $I_{\text{max}}$ and $I^{'}_{\text{max}}$ to be zero; this could degrade the sensing performance since there could be targets in the direction of users of the neighboring BSs. We discuss this further in Section \ref{result}.
We tackle the non-convex objective function using an alternating optimization algorithm. In the $i^{\text{th}}$ iteration of the algorithm, we solve the following two optimization problems alternatively.

For given $I_{i-1,\text{max}}$, $I^{'}_{i-1,\text{max}}$, $\{\mathbf{W}^{i-1,}_{mk}\}$,
\begin{align}
&\text{(P1.1.A):} \underset{\{\mathbf{W}^{i}_{mk}\},\gamma}{\text{maximize}}\,\,\,\,  \frac{u}{\mathrm{NF}_{\mathrm{R}}} \,\, {\text{FIM}^{\text{cbf}}_{mm}} +\frac{(1-u)}{\mathrm{NF}_{\mathrm{C}}} \gamma\nonumber\\
% && \text{F}_{i,i} \geq f \quad \forall i\\
& \eqref{p1.c2.1}-\eqref{sinr},\\
&\text{rank}(\mathbf{W}^{i}_{mk}) = 1; \, \mathbf{W}^{i}_{mk} \succeq \mathbf{0} \quad \forall U_{mk} \label{rank}
\end{align}
and for an obtained solution of (P1.1.A): $\gamma^{*}$ and ${\text{FIM}^{*^\text{cbf}}_{mm}}$,
\begin{align}
&\text{(P1.1.B):} \underset{\{\mathbf{W}^{i}_{mk}\},I_{\text{max}},I^{'}_{i,\text{max}}}{\text{minimize}}\,\,\,\,  I_{i,\text{max}}+I^{'}_{i,\text{max}}\nonumber\\
& {\text{FIM}^{\text{cbf}}_{mm}}\geq {\text{FIM}^{*^\text{cbf}}_{mm}} \quad \forall m\\
&\text{tr}\left (  \mathbf{Q}_{m,mk}  \mathbf{W}^{i}_{mk} \right )-\gamma^{*}\left(I^{'}_{i,\text{max}} + I_{i,\text{max}} \right) \geq \gamma^{*} \sigma^{2}_{R} \,\, \forall U_{mk},\\
& \eqref{p1.c2.1}-\eqref{intracell}, \eqref{rank}.
\end{align}
In the $i^{\text{th}}$ iteration, $\mathbf{C}_{m,\text{cbf}}$ is estimated using $\mathbf{W}^{i-1}_{mk}$. This makes the objective function an affine function of $\mathbf{R}_{x_{m}}=\sum_{k=1}^{K}\mathbf{W}_{mk}$. (P1.1.A) maximizes the weighted combination of the sensing and communication performance metrics for a given inter-cell and intra-cell interference values, whereas (P1.1.B) minimizes the interference values while guaranteeing given sensing and communication performance. Omitting the rank constraint, (P1.1.A) and (P1.1.B) are convex optimization problems solved using MATLAB's CVX solver. The overall procedure is given in Algorithm \ref{algo1}. 
\begin{algorithm}[]
\caption{CBF: precoder design}\label{algo1}
%\SetAlgoLined
\textbf{Input}: $u$, $\{\mathbf{h}_{m,nk}\}$, $\{\mathbf{G}_{mn}\} $, $I_{0,\text{max}}$, $I^{'}_{0,\text{max}}$,$\{\mathbf{W}^{0}_{mk}\}$, $i=0$\\
%   Find the minimum $n_i$ from Table \ref{table1} such that $R \geq x_{n_i} R_{p} (D^{d}_{\text{min}}, D^{u}_{\text{min}},\theta); n=n_{i-1}\times n_{i}$\\
% Solve (P1.2) for all the BSs to obtain $\{\mathbf{W}_{m,k}\}$ $\forall m ={1,2,..J}$\\
% Obtain $\{\mathbf{w}_{m,k}\}$ from $\{\mathbf{W}_{m,k}\}$ using Eigenvalue decomposition\\
% Estimate the inter-cell and intra-cell interference values to all the users: $\{I_m\}$ and $\{I^{'}_m\}$\\
\While{no convergence}
{
Determine $\mathbf{C}_{m,\text{cbf}}$ using $\{\mathbf{W}^{i}_{mk}\}$\\
$i=i+1$\\
Solve (P1.1.A) for each BS: $\gamma^{*}$ and $\{{\text{FIM}^{*^\text{cbf}}_{mm}}\}$\\
Solve (P1.1.B) to obtain $I_{i,\text{max}}$, $I^{'}_{i,\text{max}}$,$\{\mathbf{W}^{i}_{mk}\}$\\
}
\textbf{Output}:{$\{\mathbf{W}^{i}_{mk}\}$}.
\end{algorithm} 
We can obtain $\mathbf{w}_{m,k}$ from $\mathbf{W}^{i}_{mk}$ using Eigen value decomposition technique if the rank of $\mathbf{W}^{i}_{mk}\,>\,1$ 
\subsection{Coordinated Multi-point}\label{seccomp}
In a CoMP scenario, the data transmitted to a user is shared among the BSs. Let $\mathbf{X}=\left[\mathbf{X}_{1};\mathbf{X}_{2},...;\mathbf{X}_{J}\right]\in \mathcal{C}^{N \times L}$ be the concatenated symbol matrix available at each BS with $N=JN_{\mathrm{t}}$; $\mathbf{D}_{m}=\text{diag}( \mathbf{0}_{N_{\mathrm{t}}},.., \mathbf{I}_{N_\mathrm{t}},.., \mathbf{0}_{N_{\mathrm{t}}})\, \in \mathcal{C}^{N \times N}$; $\mathbf{v}^{'}_{mm}=\{\Vec{0}_{N_{\mathrm{t}}},..,\mathbf{v}_{m,m},..,\Vec{0}_{N_{\mathrm{t}}}\}\, \in \mathcal{C}^{N \times 1}$. Since each BS knows the transmitted symbol matrix of the other BSs, the ICR of \eqref{yr} aids the target's angle estimation. Consequently, the received echo at the $m^{\text{th}}$ BS is a multi-variate Gaussian random variable with mean $\boldsymbol{\mu}_{m,\text{cmp}} = \mathbf{G}^{'}_{m,m}\mathbf{D}_{m}\mathbf{X}+\sum_{n\neq m}^{J} \mathbf{G}^{'}_{n,m}\mathbf{D}_{n}\mathbf{X}$ where $\mathbf{G}^{'}_{n,m}=\alpha_{n,m}\mathbf{a}_{m,m}\mathbf{v}^{{'}^{\mathrm{T}}}_{nm}$
and the covariance matrix $\mathbf{C}_{m,\text{cmp}} =  \sigma^{2}_{\mathrm{R}}\mathbf{I}_{N_{\mathrm{r}}}$.

As an indicative example, let us take the case with $J=2$. %Using $\mathbb{\mu}_{m,\text{cmp}}$, for the first BS,
% \begin{align}
% \frac{d \mathbb{\mu}_{1,\text{cmp}}}{d \mathbf{\theta}_{11}} &= \alpha_{1,1}\dot{\mathbf{a}}_{1,1}\mathbf{v}^{T}_{1,1}\mathbf{D}_{1}\mathbf{X}+ \alpha_{1,1}{\mathbf{a}}_{1,1}\dot{\mathbf{v}}^{T}_{1,1}\mathbf{D}_{1}\mathbf{X}\\\nonumber
% &+ \alpha_{2,1}\dot{\mathbf{a}}_{1,1}\mathbf{v}^{T}_{2,1}\mathbf{D}_{2}\mathbf{X}.
% \end{align}
Adapting the CRB derivation from \cite{li2007range} to the CoMP case, for $m, n\, \in \{1,2\}$ and $\,m\neq n$, we get,
\begin{align}
& \text{tr}\left\lbrace \frac{d \boldsymbol{\mu_{m,\text{CoMP}}}^{H}}{d \mathbf{\theta}_{mm}} \mathbf{C}^{-1}_{m,\text{cmp}} \frac{d \boldsymbol{\mu_{m,\text{CoMP}}}}{d \mathbf{\theta}_{mm}} \right\rbrace \nonumber\\ &= L\alpha^{2}_{m,m}(\dot{\mathbf{a}}^{H}_{mm} \mathbf{C}^{-1}_{m,\text{cmp}} \dot{\mathbf{a}}_{mm}) \circ  (\mathbf{v}^{'^H}_{mm}\mathbf{D}_{m}\mathbf{R}^{*}_{\mathbf{X}}\mathbf{D}^{H}_{m}\mathbf{v}^{'}_{mm})\nonumber\\
    &+ L\alpha^{2}_{mm}(\dot{\mathbf{a}}^{H}_{mm} \mathbf{C}^{-1}_{m,\text{cmp}} {\mathbf{a}}_{mm}) \circ  ({\mathbf{v}}^{'^H}_{mm}\mathbf{D}_{m}\mathbf{R}^{*}_{\mathbf{X}}\mathbf{D}^{H}_{m}\dot{\mathbf{v}}^{'}_{mm})\nonumber\\
    &+ L\alpha^{2}_{mm}({\mathbf{a}}^{H}_{mm} \mathbf{C}^{-1}_{m,\text{cmp}} \dot{\mathbf{a}}_{mm}) \circ  ({{\dot{\mathbf{v}}}}^{'^H}_{mm}\mathbf{D}_{m}\mathbf{R}^{*}_{\mathbf{X}}\mathbf{D}^{H}_{m}\mathbf{v}^{'}_{mm})\nonumber\\
    &+L\alpha^{2}_{mm}({\mathbf{a}}^{H}_{mm} \mathbf{C}^{-1}_{m,\text{cmp}} {\mathbf{a}}_{mm}) \circ  ({{\dot{\mathbf{v}}}}^{'^H}_{mm}\mathbf{D}_{m}\mathbf{R}^{*}_{\mathbf{X}}\mathbf{D}^{H}_{m}{\dot{\mathbf{v}}}^{'}_{mm})\nonumber\\
     &+L\alpha_{nm}\alpha_{mm}(\dot{\mathbf{a}}^{H}_{mm} \mathbf{C}^{-1}_{m,\text{cmp}} \dot{\mathbf{a}}_{mm}) \circ  ({{\mathbf{v}}}^{'^H}_{nm}\mathbf{D}_n\mathbf{R}^{*}_{\mathbf{X}}\mathbf{D}^{H}_{m}\mathbf{v}^{'}_{mm})\nonumber\\
     &+L\alpha_{nm}\alpha_{mm}({\mathbf{a}}^{H}_{mm} \mathbf{C}^{-1}_{m,\text{cmp}} \dot{\mathbf{a}}_{mm}) \circ  ({\mathbf{v}}^{'^H}_{nm}\mathbf{D}_n\mathbf{R}^{*}_{\mathbf{X}}\mathbf{D}^{H}_{m}{\dot{\mathbf{v}}}^{'}_{mm})\nonumber\\
     &+L\alpha_{nm}\alpha_{mm}(\dot{\mathbf{a}}^{H}_{mm} \mathbf{C}^{-1}_{m,\text{cmp}} \dot{\mathbf{a}}_{mm}) \circ  ({{\mathbf{v}}}^{'^H}_{mm}\mathbf{D}_m\mathbf{R}^{*}_{\mathbf{X}}\mathbf{D}^{H}_{n}\mathbf{v}^{'}_{nm})\nonumber\\
     &+L\alpha_{nm}\alpha_{mm}(\dot{\mathbf{a}}^{H}_{mm} \mathbf{C}^{-1}_{m,\text{cmp}} {\mathbf{a}}_{mm}) \circ  ({{\dot{\mathbf{v}}}}^{'^H}_{mm}\mathbf{D}_m\mathbf{R}^{*}_{\mathbf{X}}\mathbf{D}^{H}_{n}\mathbf{v}^{'}_{nm})\nonumber\\
     &+L\alpha_{nm}\alpha_{mm}(\dot{\mathbf{a}}^{H}_{mm} \mathbf{C}^{-1}_{m,\text{cmp}}\dot{\mathbf{a}}_{mm}) \circ  ({{\mathbf{v}}}^{'^ H}_{nm}\mathbf{D}_n\mathbf{R}^{*}_{\mathbf{X}}\mathbf{D}^{H}_{n}\mathbf{v}^{'}_{nm}). \label{compfim}
\end{align}
The FIM values for the two BSs, $\text{FIM}^{\text{comp}}_{mm}$, can be obtained by substituting \eqref{compfim} in \eqref{fim}. %In the considered ComP case, all the $2K$ communication users can be considered to be served by a single virtual base station with $N$ antennas. 
Please note that, as in the CBF case, \eqref{compfim} assumes that each BS only estimates its target's angle. Let $\mathbf{h}_{k}\in \mathcal{C}^{N\times 1}$ be the channel vector from all the BSs to the $k^{\text{th}}$ user where $k=\{1,2,...,2K\}$. The corresponding optimization problem in the CoMP scenario can be formulated as,
\begin{align}
&\text{(P2):}  \underset{\{\mathbf{W}_{k}\},\eta, \gamma}{\text{maximize}}\,\,\,\,  u \,\, f +(1-u) \gamma,\nonumber\\
& \text{FIM}^{\text{comp}}_{mm} \geq f, \quad \forall m,\label{p2.c1}\\
& \frac{|\mathbf{h}^{T}_{k}\mathbf{w}_{k}|^{2}}{\sum_{l\neq k}^{2K}|\mathbf{h}^{T}_{k}\mathbf{w}_{l}|^{2}+\sigma^{2}_{\mathrm{C}}} \geq \gamma, \,\, \forall k, \label{p2.c2}\\
& \sum_{k=1}^{K}\text{tr}\left ( \mathbf{D}_{m}\mathbf{w}_{k}\mathbf{w}^{H}_{k}\mathbf{D}^{H}_{m} \right ) \leq  P_t, \forall m .\label{p2.c3}  
\end{align}
The objective function of (P2) is to maximize the weighted combination of the minimum FIM and SINR values. \eqref{p2.c2} and \eqref{p2.c3} are the SINR and per-BS power constraints, respectively. Using the SDR technique (P2) is reformulated as
\begin{align}
&\text{(P2.1):}  \underset{\{\mathbf{W}_{k}\},f, \gamma}{\text{maximize}}\,\,\,\,  u \,\, f +(1-u) \gamma,\nonumber\\
& \text{FIM}^{\text{comp}}_{m,m} \geq f, \quad \forall m,\label{p2.1.c1}\\
& \text{tr}\left (  \mathbf{Q}_{k}  \mathbf{W}_{k} \right )-\gamma\left(\sum_{l\neq k}^{2K} \text{tr}\left (  \mathbf{Q}_{k}  \mathbf{W}_{l} \right ) \right)\geq \gamma \sigma^{2}_{\mathrm{C}}, \,\, \forall k, \label{p2.1.c2}\\
& \sum_{k=1}^{K}\text{tr}\left ( \mathbf{D}_{m}\mathbf{W}_{k}\mathbf{D}^{H}_{m} \right ) \leq  P_t, \label{p2.1.c3} \\
&\mathbf{W}_{k} \succeq \mathbf{0} \, \forall k,m, \, \text{rank}(\mathbf{W}_{k}) = 1.  \label{p2.1.c4}
\end{align}
where $\mathbf{Q}_{k}=\mathbf{h}_{k}\mathbf{h}^{H}_{k}\,\in \mathcal{C}^{N \times N}$ and $\mathbf{W}_{l}= \mathbf{w}_{l}\mathbf{w}^{H}_{l}\,\in \mathcal{C}^{N \times N}$. Also, $\mathbf{R}_{\mathbf{X}}= \sum_{l=1}^{2K}\mathbf{w}_{l}  \mathbf{w}^{H}_{l}$. The SINR constraint \eqref{p2.1.c2} makes (P2.1) a non-convex problem. We adopt a similar approach of CBF to resolve this issue; \eqref{p2.1.c2} can be rewritten as, 
\begin{align}
&\sum_{l\neq k}^{2K} \text{tr}\left (  \mathbf{Q}_{k}  \mathbf{W}_{l} \right ) \leq I_{\text{max}},\label{compsdrsinr1}\\
& \text{tr}\left (  \mathbf{Q}_{k}  \mathbf{W}_{k} \right )-\gamma I_{\text{max}}\geq \gamma \sigma^{2}_{\mathrm{C}} \,\, \forall k. \label{compsdrsinr2}
\end{align}
We alternatively solve the following two optimization problems until convergence to obtain the optimal precoding vectors. For a given $I_{\text{max}}$,
\begin{align}
&\text{(P2.1.A):}  \underset{\{\mathbf{W}_{k}\},f, \gamma}{\text{maximize}}\,\,\,\,  u \,\, f +(1-u) \gamma\nonumber\\
& \eqref{p2.1.c1}, \eqref{p2.1.c3} - \eqref{compsdrsinr2}
\end{align}
For given $f^{*}$ and $\gamma^{*}$ values
\begin{align}
&\text{(P2.1.B):}  \underset{\{\mathbf{W}_{k}\}, I_{\text{max}}}{\text{minimize}}\quad I_{\text{max}}\nonumber\\
& \text{FIM}^{\text{comp}}_{m,m} \geq f^{*} \quad \forall m=\{1,2\}\\
& \text{tr}\left (  \mathbf{Q}_{k}  \mathbf{W}_{k} \right )-\gamma^{*} I_{\text{max}}\geq \gamma \sigma^{2}_{\mathrm{C}} \,\, \forall k.\\
& \eqref{p2.1.c3} - \eqref{compsdrsinr1}
\end{align}
The overall precoder design procedure is similar to Algorithm 1 without step 3. 
\section{Numerical Evaluation} \label{result}
% \begin{table}[]
% \centering
% \caption{Simulation Parameters}
% \begin{tabular}{lll}
% \hline
% Label & Definition & Value \\ \hline
% \hline
% $J$ & Number of Base station & 2\\
% $f_c$ & Channel carrier frequency & 76.5 GHz \\
% $K$ & Number of users per cell & 3\\
% $B$ & Channel bandwidth for each GN & 160 MHz \\
% $N_0$ & Noise spectral power & -174 dBm/Hz \\
% $P_t$ & Transmission Power & 30 dBm \\ 
% $N^{\mathrm{t}}_{m}\,\forall m$ & Number of Tx antennas & 16\\
% $N^{\mathrm{r}}_{m}\,\forall m$ & Number of Rx antennas & 4 \\ \hline
% \end{tabular}
% \label{table1}
% \end{table}
\begin{figure}{}
\centering
\includegraphics[width=0.85\columnwidth]{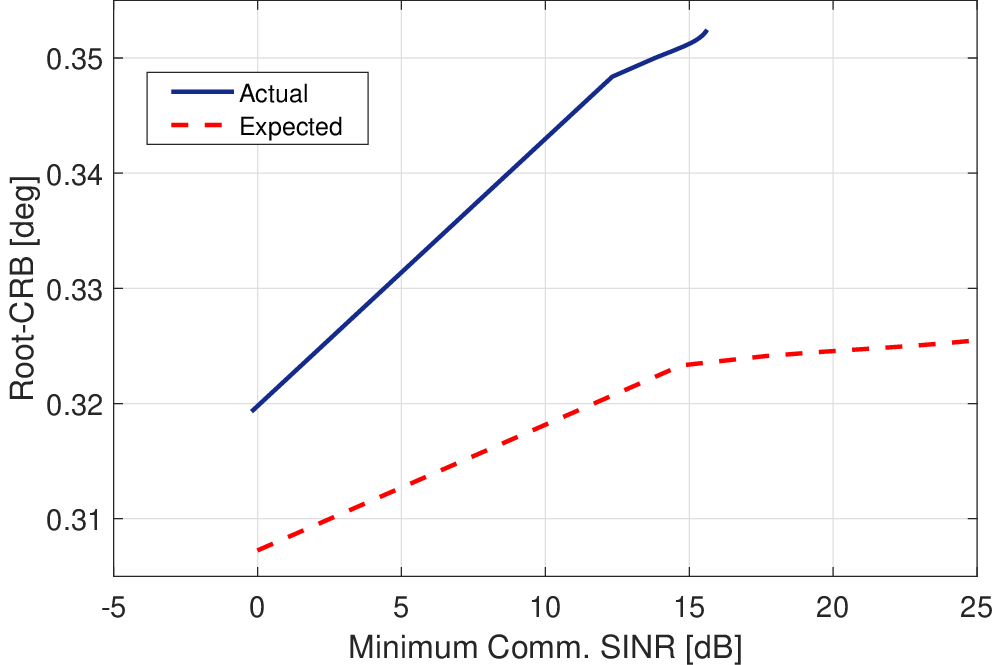}
\caption{RCRB performance gap if IC links are neglected, $N_t =6$, $N_r =4$.}\label{fig2}
\end{figure}
In this section, we summarize our main findings through numerical evaluation. The simulation parameters are $J=2$, $K=3$, $P_{t}=40$ dBm; the noise variances $\sigma^{2}_{\mathrm{C}}=\sigma^{2}_{\mathrm{R}}=0$ dBm. The targets are located at $\theta_{11}=-50^{\circ},\, \theta_{12}=60^{\circ},\,\theta_{22} = 50^{\circ},\, \theta_{21} = -60^{\circ}$. 

Fig. \ref{fig2} shows that neglecting the ICR can decrease the target angle estimation accuracy. Here, we design the precoders of the two BSs independently, neglecting the ICR and interference from the neighboring BS. We estimate the expected and actual root-CRB (RCRB) values using the obtained solution by neglecting and considering the ICR. The performance gap increases more in the high minimum communication SINR regime since the target of the neighboring BS receives more ICR power. This emphasizes the need to consider the reflections from the neighboring BSs while designing the precoders that maximize both sensing and communication performances. 

\begin{figure}{}
\centering
    % \begin{subfigure}[b]{0.49\columnwidth}
    \centering
    \includegraphics[width=0.7\columnwidth]{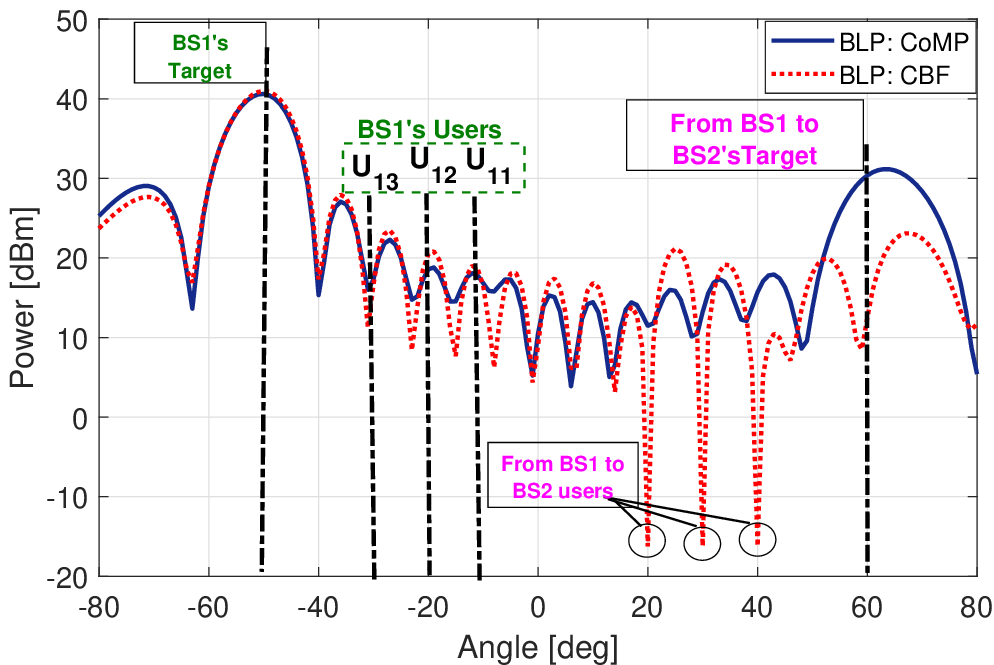}
    % \caption{CBF BS1.}\label{fig3a}
    % \end{subfigure}
    % \begin{subfigure}[b]{0.49\columnwidth}
    % \centering
    % \includegraphics[width=\columnwidth]{CoMP_final_beam_pattern_gamma_30dBm_BS1.eps}
    % \caption{CoMP BS1.}\label{fig3b}
    % \end{subfigure}
    \caption{Beampatterns for CBF and Comp, when $\gamma = 30$ dBm, $N_{\mathrm{t}}=16$, $N_{\mathrm{t}}=4$.}\label{fig3}
\end{figure}

\begin{figure}{}
\centering
\includegraphics[width=0.7\columnwidth]{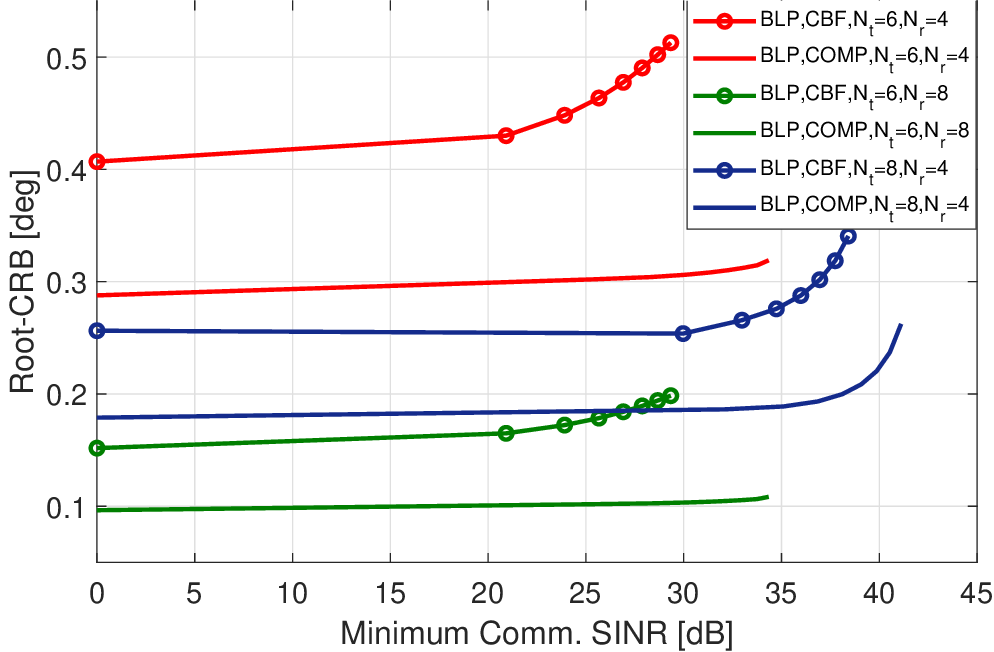}
\caption{RCRB Vs Minimum Communication SINR.}\label{fig4}
\end{figure}

For a better understanding of the beampattern towards the users and the targets, for Fig. \ref{fig3}, we consider LoS channels between the users and the BSs with $N_{\mathrm{t}}=16$ and $N_{\mathrm{r}}=4$. %The users and the target are at a distance of 25 m from the location of the corresponding BS, whereas the users and the target of the neighboring BS are at a distance of 40 m. The normalization factors $\mathrm{NF}_{\mathrm{R}}$ and $\mathrm{NF}_{\mathrm{C}}$ are obtained by solving the corresponding problem with $u=1$ and $u=0$, respectively.
%Fig. \ref{fig2} shows the effect of considering the inter-cell reflection on CRB in estimating the angle of the target of the first BS. Here we design the precoders for the users of the two BSs independently by solving the problem (32) of \cite{Liu2021}, neglecting the inter-cell reflection and interference from the neighboring BS. Using the solution, we estimate the approximate and actual root-CRB values by neglecting and considering the inter-cell reflection, respectively. The obtained values are plotted for different values of minimum communication SINR ($\Gamma$) at the users. The figure shows that neglecting the intercell reflection can decrease the target parameter estimation. The CRB gap increases more in the high minimum communication SINR regime since the target of the neighboring BS, lying in the direction of the users of the other BS, receives more inter-cell reflection power. This emphasizes the need to consider the reflections from the neighboring BSs while designing the precoders that maximize both sensing and communication performances. 
The figure shows beampatterns plotted using the precoders obtained for CBF and CoMP using Algorithm 1. In the CoMP mode, the obtained solution radiates power in both the neighboring BS's target's and users' directions since it aids the communication and the sensing performances. Conversely, the CBF solution radiates relatively less power toward the neighboring BS's target and users to minimize interference. %Ideally, we also want to keep the signal leakage towards the neighboring BS's target as minimal as possible, which requires the BS under consideration to know the direction of the neighboring BS's target, which is not feasible to obtain in practice without data sharing. %Moreover, if there is a user in the same direction as the neighboring BS's target, limiting the power radiated towards the target will result in a reduced SINR at the user.Hence, in the CBF mode, the designer has no control over the power leaked toward the neighboring BS's target.  The above two findings are reflected in Figs. \ref{fig3a}- \ref{fig3b}. 
% \begin{figure}{}
% \includegraphics[width=0.9\columnwidth]{crb_gap.jpg}
% \caption{CRB gap while neglecting inter-cell reflection.}\label{fig2}
% \end{figure}
Fig. \ref{fig4} compares the RCRB values obtained for the CBF and CoMP schemes for different numbers of antennas. % Here, the entries of the intra-cell and inter-cell channel vectors are i.i.d. Gaussian distributed with zero mean and unit variance.%As mentioned above, in the no-coordination mode, the precoders are designed by neglecting the communication and sensing links from the neighboring BSs. As seen in the figure, %there is very little gain in the CBF mode in terms of RCRB over the no-coordination scheme since the designer has no control over the power leaked toward the neighboring BS's target due to the unavailability of the angle information. 
As seen in the figure, for a given minimum communication SINR, the sensing performance in the CoMP mode outperforms the CBF mode performance because of the additional signal power received through the inter-cell reflection and communication links. 
The figure shows that in all the cases, the RCRB increases exponentially in the high-SINR regime for a given power budget since a major share of the available power is radiated toward the users to achieve the minimum communication SINR value. Moreover, the sensing performance degrades proportionally to the increase in the minimum communication SINR when the number of antennas is low. This is because the interference towards the users increases because of a high beamwidth value, demanding more power to satisfy the minimum SINR constraint. When the number of antennas increases, the beamforming gain increases, thereby reducing the required power to achieve the communication performance. This keeps the sensing performance stable in the low-SINR regime. %  as  It should be noted that the RCRB value obtained in the NC mode is lesser than the CBF mode in the high-SINR regime: this is because the NC mode neglects the inter-cell interference from the neighboring BS, thereby radiating relatively less power toward the users to guarantee a minimum SINR value, modelled considering only the intra-cell interference. The minimum SINR value observed at the user end plotted in the figure confirms this observation. 
Hence, the CoMP mode with many antennas performs best but has an additional overhead of sharing the data and channel state information among the BSs.    
\section{Conclusion}
This paper considered the effect of inter-cell reflection (ICR) and interference while designing precoders for a multi-cell MISO ISAC system operating in coordinated beamforming (CBF) and coordinated multipoint (CoMP) modes. The considered problem maximizes a weighted combination of sensing and communication performances for a given power budget. The obtained solution suggests that neglecting the inter-cell (IC) links degrades the performance, and the  performance can be improved by carefully utilizing the additional IC links.  
\section{Acknowledgement}
This work was supported in part by the Engineering and Physical Sciences Research Council under Project EP/S028455/1
%%%%%%%%%%%%%%%%%%%%%%%%%%%%%%%%%%%
%%%%% References %%%%%%%%%%%%%%%%%%
%%%%%%%%%%%%%%%%%%%%%%%%%%%%%%%%%%%
 % \begin{thebibliography}{1}
% \bibitem{CRB}
% Besson, Olivier, and Yuri I. Abramovich. "On the Fisher information matrix for multivariate elliptically contoured distributions." IEEE Signal Processing Letters 20.11 (2013): 1130-1133.
% %\bibitem{ref2}
% %Fotouhi, Azade, et al. “Survey on UAV Cellular Communications: Practical Aspects, Standardization Advancements, Regulation, and Security Challenges," \emph{IEEE Commun. Surv. and Tut. 21.4 (2019)}: 3417-3442.
% \bibitem{fancrb}
% F. Liu, Y. -F. Liu, A. Li, C. Masouros and Y. C. Eldar, "Cramér-Rao Bound Optimization for Joint Radar-Communication Beamforming," in IEEE Transactions on Signal Processing, vol. 70, pp. 240-253, 2022.
% \bibitem{DRL_power}
% Zhang, Lin, and Ying-Chang Liang. "Deep reinforcement learning for multi-agent power control in heterogeneous networks." IEEE Transactions on Wireless Communications 20.4 (2020): 2551-2564.
% \bibitem{smkay}
% Kay, Steven M. Fundamentals of statistical signal processing: estimation theory. Prentice-Hall, Inc., 1993.
% \end{thebibliography}
\bibliographystyle{IEEEtran}
\bibliography{./bibliography.bib}

\end{document}